# Comment on a recent conjectured solution of the three-dimensional Ising model

Fa Yueh Wu[1], Barry M. McCoy[2], Michael E. Fisher[3] and Lincoln Chayes[4]

[1]Department of Physics, Northeastern University, Boston, Massachusetts 02115, USA

[2]Institute for Theoretical Physics, State University of New York, Stony Brook, New York 11794-3840, USA

[3]Institute for Physical Science and Technology, University of Maryland, College Park, Maryland 20742, USA

[4]Department of Mathematics, University of California, Los Angeles, California 90059-1555, USA

Abstract

In a recent paper published in the Philosophical Magazine [Z.-D. Zhang, *Phil. Mag.* **87**, 5309-5419 (2007)], the author advances a conjectured solution for various properties of the three-dimensional Ising model. Here we disprove the conjecture and point out the flaws in the arguments leading to the conjectured expressions.

The Ising model [1] is a well-known and well-studied model of magnetism. Because of its apparent simplicity, the model has attracted the concerted attention of physicists for over 80 years. Ising solved the model in one dimension in 1925. In 1944 Onsager [2] obtained the exact free energy of the two-dimensional (2D) model in zero field, and in 1952 Yang [3] presented a computation of the spontaneous magnetization. But the three-dimensional (3D) model has withstood challenges and remains to this date an outstanding unsolved problem.

In a recent 111-page paper published in *Philosophical Magazine*, Zhang [4] advanced conjectured expressions for the free energy and spontaneous magnetization of the 3D model. Here we show that the conjectures are false.

Zhang [4] considered the nearest-neighbor 3D Ising model on the simple cubic lattice: see his Eqs. (1) and (2) and associated text where the notations are established with, specifically, three coupling constants $K = J/k_BT$, $K'$, and $K''$. Arguments leading to the conjectured solutions are roughly as follows:[1]

The author presents an expression, Eq. (49), in the form of a 4-fold integral [which reduces to Eq. (74) in the isotropic case] as the exact free energy. But this expression contains yet-to-be-determined, unknown weight functions $w_x, w_y, w_z$. The argument next jumps to Appendix A where the author sets $w_x = 1$ and expands $w_y = w_z$ in the form of a square root of a series: see Eq. (A2). Also in the Appendix, the author demonstrates that the expansion coefficients of the first 11 terms of the series can be fitted, as shown in (A2), to ensure that (74) reproduces the known 11 terms of the exact high-temperature expansion of the free energy obtained by Guttmann and Enting [5]; see also line 1, p. 5326.

---

[1] Key assumptions made in [4] are not presented in a logical sequence but are often hidden in inconspicuous places, making it difficult for a reader to see what is really going on.

Almost as if in "fine print," the author then sets $w_y = w_z = 0$ — see line 7 on p. 5326 just before Eq. (50) — and uses the resulting form of (49) as the conjectured solution of the free energy throughout the ensuing analysis where conclusions on the critical point, etc., are drawn. The reason given for taking $w_y = w_z = 0$ is what the author calls "Ansatz 1" in Appendix A (p. 5399). Under this ansatz the author argues (lines 7-9, p. 5400), the series inside the square root would become negative making $w_y$ and $w_z$ imaginary. Since imaginary quantities are "physically not meaningful", $w_y$ and $w_z$ "are always equal to zero" (p. 5400).

It must be emphasized that this argument for choosing the weights $w_y = w_z = 0$, is deeply flawed. Indeed, in light of the fitting of the series in (A2) to reproduce the known high-temperature expansions, one knows that the choice $w_y = w_z = 0$ will *not* reproduce the exact high-temperature expansions. Hence the resulting expressions (49) and (74) *cannot* be the true solution of the free energy. By the same token, the "putatively determined" critical point relations (see the *Abstract*, etc.) carry no credence.

For the spontaneous magnetization the author presents the expression (99) [reducing to (102) in the isotropic case] as the exact solution. But this expression is again obtained by using the flawed choice of $w_y = w_z = 0$ [see 4 lines below Eq. (86), p. 5339]. This mistaken procedure leads to a critical exponent $\beta = 3/8$ for the magnetization of the 3D model. But it also gives the same exponent 3/8 for the 2D model — since Eq.(99) reduces to 2D by setting $K'' = 0$ or $x_4 = 1$. This is clearly wrong since we know from the exact solution of Yang [3] that the 2D exponent is 1/8. Moreover, the expansion of the expression (102), namely, $1 - 6x^8 - 12x^{10} - 18x^{12} - \cdots$ [see Eq.(103)], fails to agree with the exact low-temperature expansion of the spontaneous magnetization of the simple cubic lattice [6] which is $1 - 2x^6 - 12x^{10} + 14x^{12} - \cdots$.

A cardinal, golden rule for verifying the validity of any proposed exact solution is that it must yield, term by term, the correct high and low temperature expansions. Indeed, in many cases including, in particular, the case of the three-dimensional Ising ferromagnet, this is the subject of a mathematical theorem (see Sinai [7]). Since the author clearly realizes that his conjectured expressions fail in this test, he has assembled a variety of reasons to justify the failure. He states that the test works in $d = 2$ dimensions because "in the 2D case, we are extremely lucky because both the high- and low-temperature expansions are exact and convergent" (Sec. 8.2.3, p. 5382, 13th line in second paragraph). To explain the failure of the conjectured free energy, for example, the author argues that the known exact high-temperature expansion holds only "at/near" infinite temperature — see line 1, p. 5331 and 4 lines below (A13), p. 5406 — and thus for finite temperatures one must use the weights $w_y = w_z = 0$. This argument of arbitrarily dividing "at/near infinite" and "finite temperatures" is flawed. Indeed, the suggestion contradicts general rigorous results establishing the finite radius of convergence of the high-$T$ and low-$T$ expansions and their exact representation of the thermodynamic limit for all $d \geq 2$ [7].

To explain the incorrect prediction of $\beta = 3/8$ for the 2D spontaneous magnetization, the author argues in Sec. 4.2 that there exists a certain region in the interaction parameter space where the exponent $\beta$ crosses over from the 3D value 3/8 to the 2D value 1/8. This suggestion is contrary to well-established understanding of critical phenomena and crossover behavior and is, thus, implausible. To patch up the disagreement of (102) with the exact low-temperature expansion, the author states as his opinion "that the requirement, where the exact expression must be equal, term by term, to the so-called *exact* low-temperature expansion has, for a long time, reflected a pious hope": see the first paragraph of Sec. 8.2.2., p. 5377. This opinion, as noted above, contradicts a host of long established rigorous results for Ising and more general models [7].

In summary, Zhang's suggestion that the free energy be expressed as a 4-fold integral has not produced a solution to the 3D Ising model. Specifically, the conjectured expressions (74) and (99), in which the crucial temperature-dependent weights $w_y$ and $w_z$ have been set to zero, cannot be exact solutions. Furthermore, the arguments advanced for this step are unsupported and hence carry no conviction. In conclusion, the various conjectured relations for the value of $T_c$, for critical exponents, etc., including others not discussed in this note (such as the true range of correlation in Sec. 5.4) are false.